\documentclass[aps, pra, twocolumn, superscriptaddress, amsmath, tightenlines, longbibliography]{revtex4-2}

\usepackage{color}

\usepackage{amssymb}
\usepackage{amsmath}
\usepackage{dcolumn}
\usepackage{graphicx}
\usepackage{mathrsfs}
\usepackage{appendix}
\usepackage{graphicx}
\usepackage{booktabs}
\usepackage{colortbl}
\usepackage{units}

\definecolor{Dred}{RGB}{190,0,0}

\def \hide#1{}

\newcommand{\ket}[1]{\mbox{$|#1\rangle$}}

\usepackage{url}
\usepackage[colorlinks]{hyperref}
\hypersetup{%
	plainpages=true,
	breaklinks=true,       
	hypertexnames=false,  
	pageanchor=true,
	colorlinks=true,
	linkcolor={blue},
	citecolor={red},
	urlcolor={blue},
	anchorcolor={black}
}

\begin{document}

\title{Supercorrelated decay in a quasiperiodic nonlinear waveguide: From Markovian to non-Markovian transitions}

\author{Jia-Qi Li}
\author{Tian-Yu Zhou}
\author{Xin Wang}
\email{wangxin.phy@xjtu.edu.cn}
\affiliation{Institute of Theoretical Physics, School of Physics, Xi’an Jiaotong University, Xi’an 710049, People’s Republic of China}

\date{\today}

\begin{abstract}

Mobility edges (MEs) are critical boundaries in disordered quantum systems that separate localized from extended states, significantly affecting transport properties and phase transitions. Although MEs are well-understood in single-photon systems, their manifestation in many-body contexts remains an active area of research. In this work, we investigate a one-dimensional Bose-Hubbard chain with a quasiperiodic potential modulating photon-photon interactions, effectively creating a mosaic lattice. We identify MEs for doublon states (i.e, bound photon pairs resulting from strong interactions) within the two-photon subspace. Our analytical solutions and numerical simulations confirm the existence of these MEs, extending single-photon MEs theories to the two-photon regime. Additionally, we analyze the dynamics of two emitters coupled to the waveguide, enabling the emission of supercorrelated photon pairs into the waveguide. Our findings reveal that coupling to extended states results in Markovian dynamics, characterized by exponentially supercorrelated decay, while coupling to localized states gives rise to non-Markovian dynamics, marked by suppressed decay and persistent oscillations. Here, a transition from Markovian to non-Markovian behavior occurs around the MEs of the doublons. Finally, we propose a feasible experimental implementation using superconducting circuits, providing a platform to observe the interplay between interactions and disorder in quantum systems.
\end{abstract}

\maketitle

\section{Introduction}

Anderson localization~\cite{Anderson1958}, found in 1958, is fundamental for understanding disordered quantum systems by demonstrating how disorder inhibits wave diffusion. In one- and two-dimensional (1D, 2D) systems, localization occurs regardless of disorder strength and has been observed in ultracold atoms, optical lattices, and photonic crystals~\cite{Naether2012,Schwartz2007,Sttzer2012,Titum2015,GuzmanSilva2020,Jovi2011,Lahini2008,Martin2011}. In three-dimensional (3D) systems, localization arises only when disorder exceeds a critical threshold, leading to MEs at energy $ E_c $ that separate extended and localized states. This results in the coexistence of both states, phase transitions between conducting and insulating phases~\cite{Bulka1987,Evers2008,Lugan2011,Mott1968}, and critical scaling near MEs~\cite{Slevin1999,Vasquez2008}. Recent studies also show that ME-like transitions can emerge from dissipation and decoherence~\cite{Liu2024,Longhi2024}, broadening the understanding of MEs.

Interestingly, quasiperiodic potentials~\cite{Biddle2010,Boers2007,DasSarma1988,Ganeshan2015,Liu2022,Xu2020,Yao2019,Wang12020}, acting as deterministic analogs to random disorder, can induce localization effects and exhibit MEs even in low-dimensional systems. In these systems, quasiperiodic potentials host both localized and extended states, with MEs defining the transitions between them. Furthermore, quasiperiodic systems can support a third type of state known as critical states~\cite{Deng2019,Wang2022,Zhou2023}, which possess properties intermediate between localized and extended states. This broadens the concept of MEs to include boundaries between critical states and both localized and extended states~\cite{Han1994,Liu2015,Wang2016}, garnering increasing research interest. This nuanced classification of eigenstates creates a rich landscape that significantly influences coupled quantum systems. When coupling an emitter to a waveguide, the waveguide's eigenstates critically determine the emitter’s dynamical behavior. Specifically, coupling to extended states results in Markovian dynamics characterized by exponential decay, whereas coupling to localized states induces non-Markovian dynamics~\cite{deVega2017,Gaikwa2024,Liu2011,Lombardo2014,Tufarelli2014}, where energy decay is suppressed and oscillates near a higher energy level. These observations highlight the pivotal role of quantum correlations in shaping the dynamics of coupled emitter–waveguide systems. This raises the question of whether similar phenomena arise when an emitter is coupled to a bath with MEs, a largely unexplored scenario that warrants further investigation.

Additionally, MEs are not only manifest in single-photon systems, but also emerge in multi-photon nonlinear regime, where photon-photon interactions lead to intriguing many-body phenomena~\cite{Bordia2017,Iyer2013,Sierant2017,Sierant2022,Bin2020}. With the nonlinear interaction, the intriguing multi-photon state emerge in the system, such as ``doublon'', where two photons are bunched together to transport along the waveguide\cite{daVeiga2002,Mahajan2006,Winkler2006,Piil2007,Valiente2008,Salerno2020,Azcona2021}. Moreover, due to this spatial bound property, the distance between of emitters must reside in the correlation length of the doublon, in which collective \textit{supercorrelated radiance emerges}, where two excited emitters jointly radiant and excite the doublon state\cite{Wang22020,Talukdar2022,Wangxin2024}. While MEs in many-body systems have been extensively studied, those in photonic bound states remain largely unexplored. This represents a critical gap in understanding how MEs influence doublon dynamics in 
these frameworks.

In this work, we investigate MEs within the two-photon subspace by introducing a one-dimensional quasiperiodic potential to tune the strength of photon-photon interactions, focusing specifically on doublon bands in a Bose-Hubbard chain. We analytically derive the form of MEs and further examine their impact on the radiative dynamics of two-photon states. Through numerical simulations, we reveal that MEs profoundly influence the dynamics of radiance, with variations in the quasiperiodic potential strength driving the evolution from Markovian to non-Markovian regimes. Finally, we propose a potential experimental realization of these phenomena using transmon chains with feasible parameters.

\section{The Hamiltonian of the entire system}

\begin{figure}
	\includegraphics[width=\linewidth]{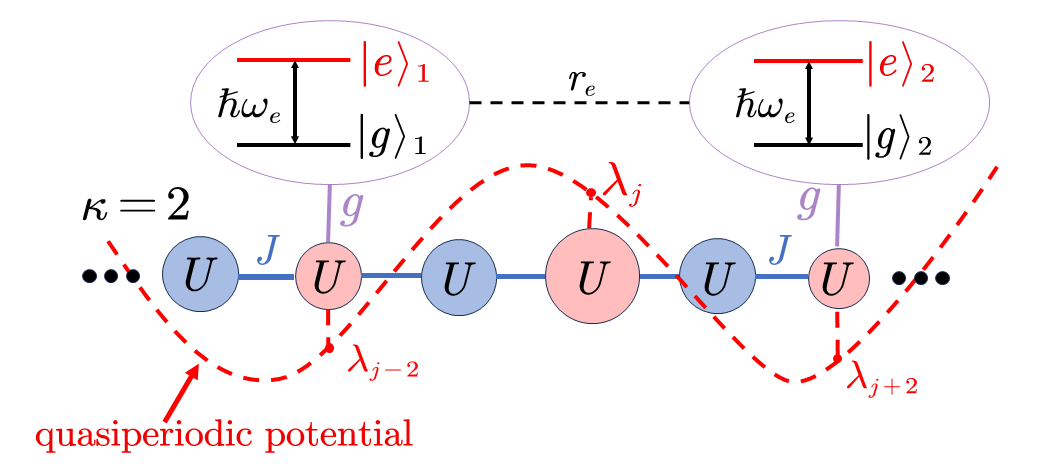}
	\caption{Schematic of the system Hamiltonian. The red (blue) spheres represent lattice sites modulated (unmodulated) by a quasiperiodic potential, with $\kappa = 2$ in the diagram. The red dashed lines illustrate the strength of the quasiperiodic potential, while the varying sizes of the red spheres reflect the modulation intensity. The nonlinear local potential $U$ characterizes the effective photon-photon interaction; the $U$ term of the Eq.~\eqref{BH}, only plays a role for $N \geq 2$. The solid dark blue lines represent the hopping constant between waveguide sites. The two two-level emitters, each with the same frequency $\omega_e$, have energy levels $|g\rangle_i$ and $|e\rangle_i$ ($i = 1$ or $2$). These emitters are coupled to cavities $n_1$ and $n_2$ with a coupling strength $g$, where $n_1$ and $n_2$ are fixed, and $r_e = |n_1 - n_2|$. In this work, we focus on the specific case of $r_e = 0$, meaning that the two emitters are coupled at the same position.} 
	\label{fig1m} 
\end{figure}

We consider a system of two-level emitters coupled to a waveguide, where the ground and excited states of the emitters are denoted as $|g\rangle_i$ and $|e\rangle_i$, respectively (see Fig.~\ref{fig1m}
). The total Hamiltonian of the system is expressed as  
\begin{equation}  
	H = H_w + \frac{\omega_e}{2} \sum_{i=1}^N \sigma_i^z + g \sum_{i=1}^N \left( a_{n_i} \sigma_i^+ + a_{n_i}^\dagger \sigma_i^- \right),  
	\label{H_total}  
\end{equation}  
where $H_w$ represents the Hamiltonian of the nonlinear waveguide, the second term describes the two-level emitters, and the third term accounts for their coupling to the waveguide. 

In this formulation, $ \sigma_i^{\pm} $ represent the raising and lowering operators for the $i$-th two-level emitter, while $ \sigma_i^{z} $ denotes the Pauli-$z$ operator. Each emitter is coupled to the waveguide at site $n_i$, where $a_{n_i}^\dagger$ and $a_{n_i}$ are the creation and annihilation operators for excitations at this site, respectively. The emitter transition frequency is given by $\omega_e$, and the coupling strength $g$ quantifies the interaction between the emitters and the waveguide. For simplicity, we assume that both emitters are coupled to the same site, i.e., $r_e = |n_1 - n_2| = 0$.

In this work, we describe the waveguide as a Bose-Hubbard chain modulated by a one-dimensional quasiperiodic potential. The Hamiltonian governing the waveguide is expressed as
\begin{equation}
	H_w = H_{\text{BH}} + H_{\lambda},
	\label{H_waveguide}
\end{equation}
where $ H_{\text{BH}} $ describes the Bose-Hubbard chain, given by
\begin{equation}
	H_{\text{BH}} = H_0 + H_U,
	\label{BH}
\end{equation}
with
\begin{align}
	H_0 &= \sum_j \omega_c a_j^\dagger a_j - J \left( a_j^\dagger a_{j+1} + \text{H.c.} \right), \\
	H_U &= \frac{U}{2} \sum_j a_j^\dagger a_j^\dagger a_j a_j, \label{Hu}
\end{align}
and $ H_{\lambda} $, the quasiperiodic potential, which is used to tune the photon-photon interactions strength $ U $, is expressed as  
\begin{equation}
	H_{\lambda} = \sum_j \lambda_j a_j^\dagger a_j^\dagger a_j a_j.
\end{equation}
The coefficient $\lambda_j$ is defined as
\begin{equation}
	\lambda_j = \left\{
	\begin{array}{ll}
		\lambda \cos [2\pi (\omega j + \theta)], & \quad j = m\kappa, \\
		0, & \quad \text{otherwise},
	\end{array} \right.
	\label{eq:lambda}
\end{equation}
where $ a_j $ ($ a_j^\dagger $) denotes the annihilation (creation) operator for bosons at site $ j $, $ \omega_c $ is the central frequency, $ J $ represents the nearest-neighbor hopping rate, and $ U $ is the nonlinear local potential characterizing the strength of photon-photon interactions. The quasiperiodic potential $\lambda_j$ in these models exhibits a mosaic structure, where $\kappa$ is an integer defining the periodicity of the pattern. The parameters $\lambda$ and $\theta$ denote the amplitude and phase offset of the quasiperiodic potential, respectively. 

The terms $ H_0 $ and $ H_U $ in $ H_{\text{BH}} $ govern the tight-binding dynamics and the nonlinear interaction of the waveguide, respectively. These interactions give rise to two distinct classes of wavefunctions: scattering states and bound states (doublons). The inclusion of $ H_{\lambda} $ breaks the duality symmetry of the waveguide, resulting in the emergence of MEs, which delineate the transition between localized and extended states in the spectrum. By varying the strength of $ \lambda $, we can tune the energy structure, leading to different radiance profiles of the emitter's energy, such as exponentially supercorrelated decay or suppressed behavior.

This framework provides the foundation for exploring the interplay between localization phenomena and the dynamics of coupled emitter–waveguide systems. A detailed derivation and further analysis of these physical quantities will be presented in the subsequent sections.

\begin{figure}
	\includegraphics[width=\linewidth]{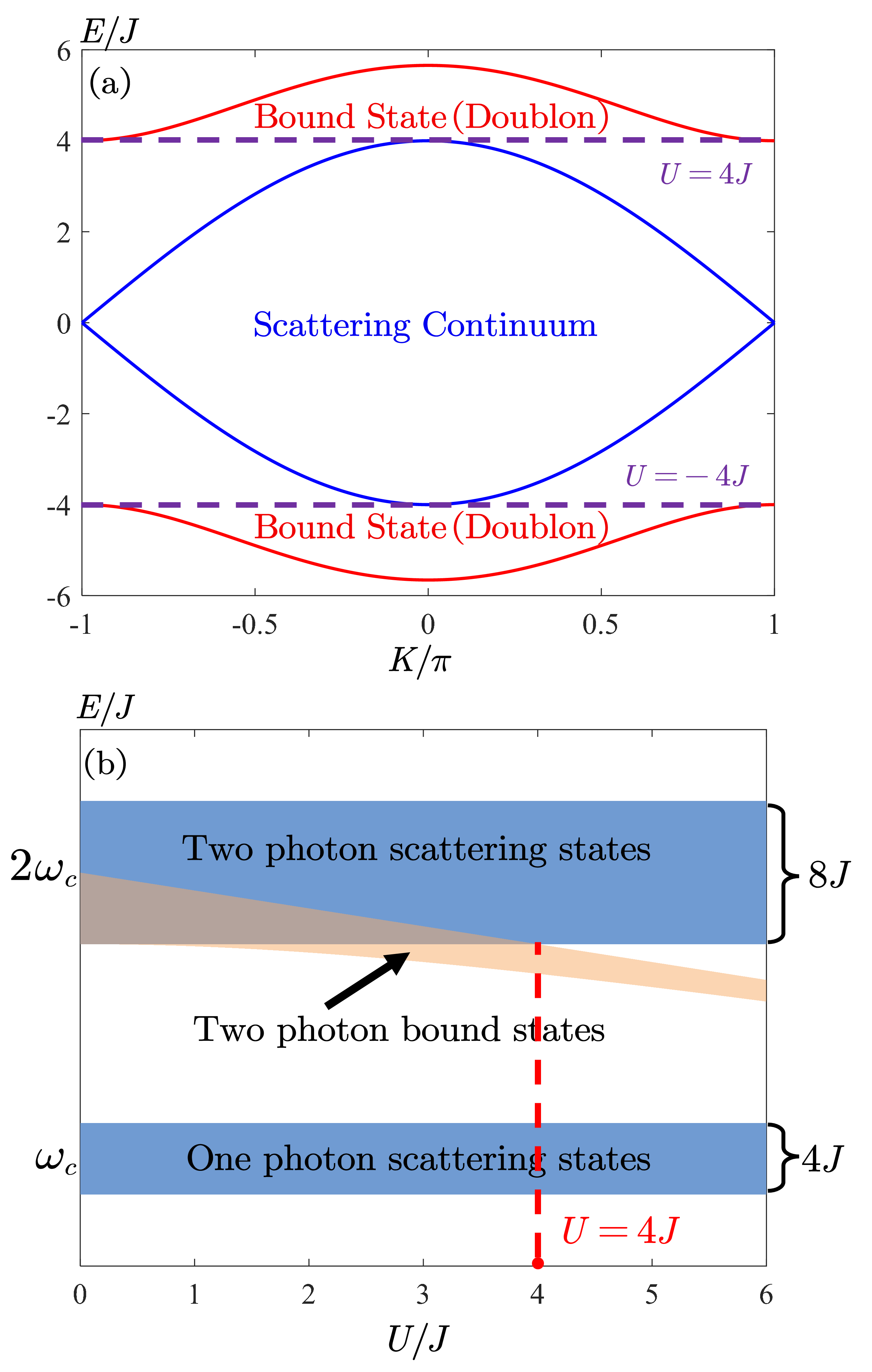} 
	\caption{(a) Schematic of the energy bands in the two-photon subspace, where the red and blue solid lines represent the bound states Eq.~\eqref{eqEB} and scattering states Eq.~\eqref{eqES}, respectively. The purple dashed line indicates the threshold where the bound and scattering states no longer overlap. The parameters used here are $U = \pm 4J$ and $\omega_c = 0$. (b) Variation of the bandwidths of the bound and scattering states with the nonlinear potential $ U $ in two-photon subspace. The bandwidth of the scattering states remains unchanged as $ U $ varies, while the bandwidth of the bound states decreases with increasing $ U $, eventually approaching a flat band.} 
	\label{fig2m}  
\end{figure}

\section{Mobility edges in two-photon subspace}

In this section, we investigate the emergence of MEs within the two-photon subspace of a one-dimensional Bose-Hubbard chain. We begin by examining the doublon states in the absence of a quasiperiodic potential, establishing the fundamental characteristics of bound photon pairs. Subsequently, we derive the effective hopping amplitude and Hamiltonian of the bound pair chain, which is essential for understanding the transport properties and dynamics of these doublon states. Finally, we explore the MEs for the bound pair chain, identifying the critical parameters that delineate the boundary between localized and extended two-photon states.

\subsection{Doublon state without quasiperiodic potential }

A striking result of solving the Bose–Hubbard Hamiltonian Eq.~\eqref{BH} is the prediction of stable repulsively bound photon pairs, known as doublons~\cite{daVeiga2002,Mahajan2006,Piil2007,Valiente2008,Winkler2006,Azcona2021,Salerno2020,Fisher1989,Jaksch2005}. The solution reveals two distinct types of eigenstates: the scattering band and the bound band. The detailed derivation and intermediate steps are provided in Appendix~\ref{appendix1}, and here we summarize the main results.   

The scattering states are unbound, allowing photons to move independently within the waveguide. The two-photon scattering spectrum has a bandwidth $ E_S = 8J $, twice that of the single-photon spectrum, with the energy band given by:
\begin{equation}
	E_S = 2 \omega_c - 4J \cos\left( \frac{K}{2} \right) \cos(k).
	\label{eqES}
\end{equation}
Here, $ K $ denotes the total momentum of the two-photon system, and $ k $ represents the relative momentum between the photons.

In contrast, the bound states, induced by the nonlinear potential $U$, form a quasiparticle known as a "Doublon". In this state, the two photons exhibit strong correlations and behave as a single entity. The energy bands for the doublon are:  
\begin{equation}
	E_{\pm}^{B} = 2 \omega_c \pm \sqrt{U^2 + \left[ 4J \cos\left( \frac{K}{2} \right) \right]^2}.
	\label{eqEB}
\end{equation}  

As shown in Fig.~\ref{fig2m}(a), the energy bands $ E_{\pm}^{B} $ (red lines) and $ E_S $ (blue lines) illustrate the distinction between these states. For $ U > 0 $ ($ U < 0 $), the doublon states lie above (below) the scattering states, indicating repulsive (attractive) interactions.

For the convenient description of two-photon subspace, we apply the relative and center-of-mass coordinate, i.e., $r=m-n$ and $x_c=(m+n)/2$. The doublon wave function has a form~\cite{Winkler2006,Piil2007}
\begin{gather}
\Psi _{\mathrm{D}}\propto \exp \left( iKx_c \right) \exp \left( -|r|/L_c \right).
\end{gather}
For $r$ direction, the wavefunction exponentially decay and $L_c$ is the correlated distance. The two photons of doublon are bound within the regimes $L_c$. However, for $x_c$ direction, the wavefunction exhibits an extended state, spreading across the entire waveguide. The two photons are bunched together to transport along the waveguide.

\subsection{The effective hopping amplitude and Hamiltonian of the bound pair chain}

We show the effect of the nonlinear potential $ U $ on the two-photon energy band. Figure.~\ref{fig2m}(b) presents the energy band structures as a function of $ U $. The bandwidths of the single- and two-photon scattering states remain unchanged versus $ U $. However, as the strength of $ U $ increases, the bandwidth of the bound states decreases, indicating that the two photons are more tightly bound, with a reduced energy range for free motion. In the limit of strong interaction $ U \gg J $, the two photons become tightly bound within the potential well, inhibiting any transitions and causing the photon wavepacket to localize at a specific lattice site. 

\begin{figure}
	\includegraphics[width=\linewidth]{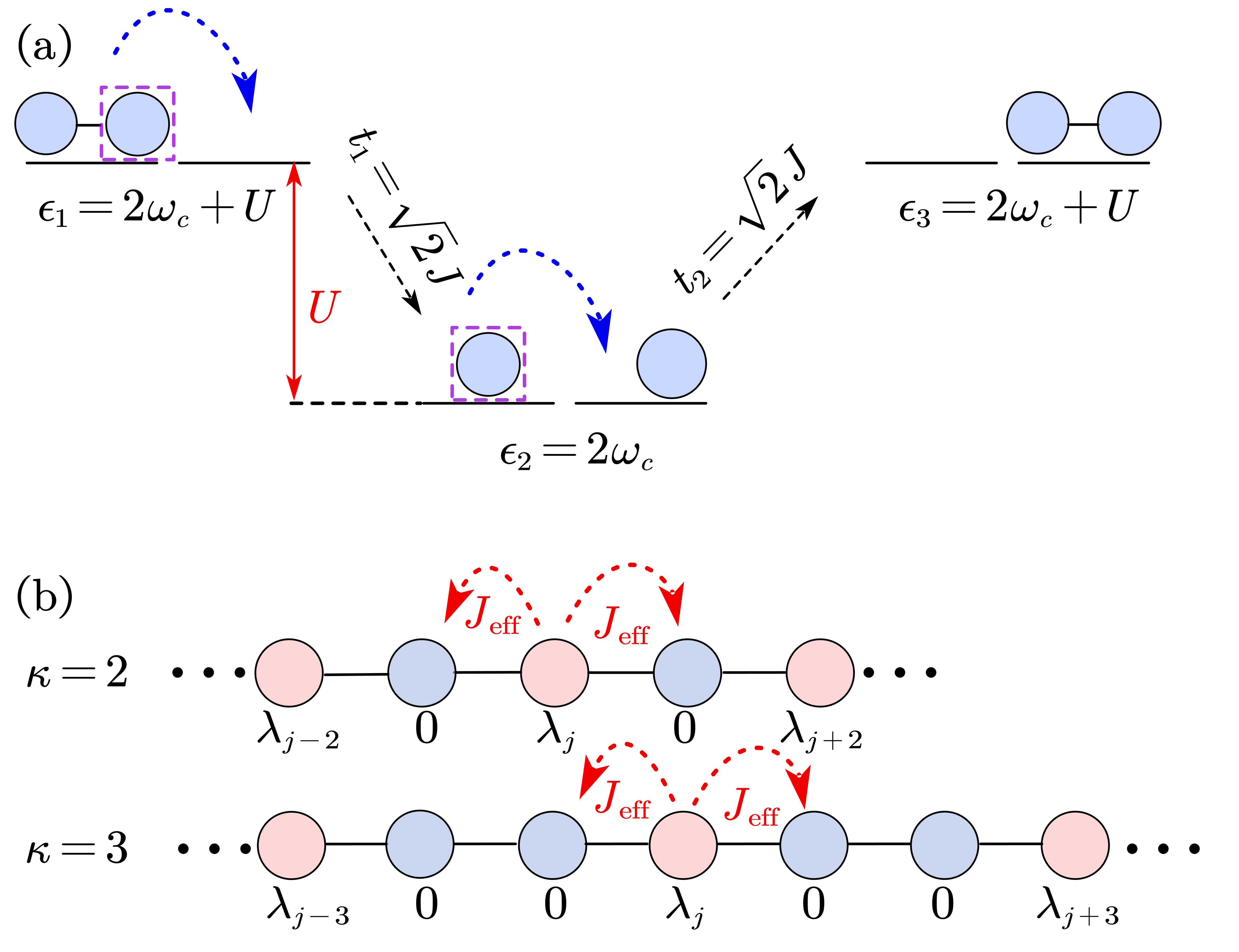}  
	\caption{(a) The second-order hopping process of the doublon pair, where $\epsilon_1$, $\epsilon_2$, and $\epsilon_3$ represent the initial, intermediate, and final states of the hopping process, respectively. (b) The 1D quasiperiodic mosaic model for $\kappa = 2$ and $\kappa = 3$ is depicted. Red spheres represent waveguide sites modulated by the quasiperiodic potential, while blue spheres indicate unmodulated sites. The black solid lines denote the effective hopping amplitude $J_\text{eff}$, and $\lambda_j$ represents the quasiperiodic mosaic potential.} 
	\label{fig3m}  
\end{figure}

To describe this phenomenon and focus on two-photon subspace, we introduce the state $|m,n\rangle$, where the two photons are localized at sites $ m $ and $ n $, respectively. Due to the indistinguishability of bosons, the transition rate between $|m,m\rangle$, in which both photons occupy the same site, and the accessible transition state $|m,m\pm1\rangle$, where one photon hops to a nearest-neighbor site, is given by
\begin{gather}
Ja_{m\pm 1}^{\dagger}a_m|m,m\rangle =\sqrt{2}J|m,m\pm 1\rangle.
\end{gather}
However, due to the nonlinear potential $U$, the state $|m,m\rangle$ exhibits an energy offset $U$ relative to states $|m,n\rangle$ ($m\ne n$) where two photons are spatially separated~\cite{Winkler2006,Piil2007}, as shown in Fig.~\ref{fig3m}(a). In the limit $U\gg J$, the state $|m,m\rangle$ is far-detuned from the accessible states $|m,m\pm1\rangle$, which implies that the direct transition process $|m,m\rangle \leftrightarrow | m,m\pm1\rangle$ is strongly suppressed. In contrast, for states $|m,n\rangle$ ($m\ne n$), there is no energy offset, and the two photons can freely hop to nearest-neighbor sites, i.e., $|m,n\rangle \leftrightarrow |m\pm 1,n\pm 1\rangle$, provided the photons are not adjacent $m\ne n\pm1$. Although the direct transition $|m,m\rangle\rightarrow|m,m+1\rangle$ is suppressed, it can serve as an \textit{intermediate virtual process} to mediate higher-order processes, $$|m,m\rangle\rightarrow|m,m+1\rangle\rightarrow|m+1,m+1\rangle,$$ as shown in Fig.~\ref{fig3m}(a). Therefore, the two photons can be treated as a quasiparticle or boson stacks \cite{Mansikkam2022} to transport along the waveguide with an effective hopping strength $J_{\mathrm{eff}}$. Using second-order perturbation theory~\cite{Mansikkam2022,Pinto2009,Gorlach2017}, the effective transition rate is derived as: 
\begin{equation}
J_{\text{eff}}\approx \frac{(\sqrt{2}J)^2}{\sqrt{U^2 + 8J^2}},
\label{Jeff}
\end{equation}
where $(\sqrt{2}J)^2$ represents the product of hopping amplitudes associated with the second-order process, and the denominator corresponds to the energy difference between the intermediate state $|m,m+1\rangle$ and the initial $|m,m\rangle$ or final $|m+1,m+1\rangle$ states. As $U$ increases, the energy offset grows larger, leading to a reduction in the transfer efficiency of quasiparticle. Additionally, the energy spectrum of the doublon state narrows, as shown in Fig.~\ref{fig2m}. We refer interesting readers to Ref.~\cite{Mansikkam2022}, where the authors provide a detailed derivation of the effective transition for the $N$-photons case, extending beyond the doublon scenario.

If the system is initially prepared with only two-photon (doublon) states populated, it can be effectively described by a "Doublon Hamiltonian" in the limit $ U \gg J $~\cite{Petrosyan2007,ValienteP2008,Wang2008}. By defining the number operator in the single-photon subspace as $ n_j = a_{j}^{\dagger}a_j $, the effective Hamiltonian can be derived to second order in $ J/U $, and involves the doublon creation operator,
\begin{equation}
	D_{j}^{\dagger}=(a_{j}^{\dagger})^2\left[ \frac{1}{2\sqrt{(n_j+1)}} \right], 
\end{equation}
and the annihilation operator,
\begin{equation}
	D_j=\left[ \frac{1}{2\sqrt{(n_j+1)}} \right] (a_j)^2.
\end{equation}
These operators obey the standard bosonic commutation relations $ [D_j, D_i^\dagger] = \delta_{ji} $ and $ [D_j, D_i] = [D_j^\dagger, D_i^\dagger] = 0 $. The doublon number operator at site $ j $ is then given by $ m_j = D_j^\dagger D_j = n_j / 2 $, and it is easy to verify by induction that 
\begin{equation}
	|m_j\rangle = \frac{1}{\sqrt{m!}} (D_j^\dagger)^m |0\rangle.	
\end{equation}
After calculation, the effective Hamiltonian for a  doublon pair is then given by~\cite{Petrosyan2007,ValienteP2008,Wang2008}: 
\begin{equation}
	H_{\text{eff}}\! =\! \sum_j\! \left[ (2\omega_c \! + \! U \! - \! 2J_{\text{eff}}) m_j \!-\! J_{\text{eff}}\! \left( D_j^\dagger \! D_{j+1} \! + \!\text{H.c.} \right)\! \right]\!.
	\label{Heff}
\end{equation}
Therefore, we obtain the effective Hamiltonian $ H_{\text{eff}} $, which describes the dynamics of doublon states in the chain, treating each doublon as a composite bosonic object, while ignoring the scattering state. The first term $ (2\omega_c+U - 2J_{\text{eff}}) m_j $ represents the internal energy of each doublon, combining the strong on-site interaction $ U $ and the effective hopping amplitude $ J_{\text{eff}} $, reflecting the energy cost of placing a doublon at a site. The second term $ - J_{\text{eff}} \left( D_j^\dagger D_{j+1} + \text{H.c.} \right) $ describes the doublon hopping between adjacent sites, with $ J_{\text{eff}} $ being the  effective hopping amplitude that is renormalized by the interactions.

From the expression for $J_{\text{eff}}$, it is evident that when $U$ is large, $J_{\text{eff}}$ is significantly reduced, indicating that the movement of the doublon is strongly suppressed. Conversely, when $U$ is small, $J_{\text{eff}}$ approaches $J$, allowing the doublon to move more freely.
Particularly, when $U < 4J$, there is an overlap between the scattering and bound states, leading to competition between the two modes in photon distribution. However, when $U \geq 4J$, this overlap disappears, allowing for the isolation of pure scattering or bound eigenstates by adjusting the value of $U$. Since this study primarily focuses on bound states, we will consider cases where $U \geq 4J$ in the subsequent analysis.

\subsection{Mobility Edges for the bound pair chain}

In this subsection, we discuss the impact of the quasiperiodic potential on doublon states. By introducing a quasiperiodic potential to modulate the nonlinear local potential $ U $ in the Bose–Hubbard chain~\cite{Valiente2008}, the system is transformed into a mosaic lattice. Specifically, we consider a system composed of $\kappa$ nearest-neighbor lattice sites grouped into $ M $ sets, where $ m = 1, 2, \dots, M $. Consequently, the total system size is $ L = \kappa M $~\cite{Wang12020}. It is evident that when $ \kappa = 1 $, the model becomes trivial. In the single-photon subspace, this further reduces to the Aubry-André-Harper model~\cite{Aubry1980}. However, for $ \kappa \neq 1 $, the duality symmetry of the chain is broken, which opens up the possibility of investigating the existence of MEs. The quasiperiodic mosaic models for $ \kappa = 2 $ and $ \kappa = 3 $ are shown in Fig.~\ref{fig3m}(b), and other cases follow a similar pattern.

\begin{figure}
	\includegraphics[width=\linewidth]{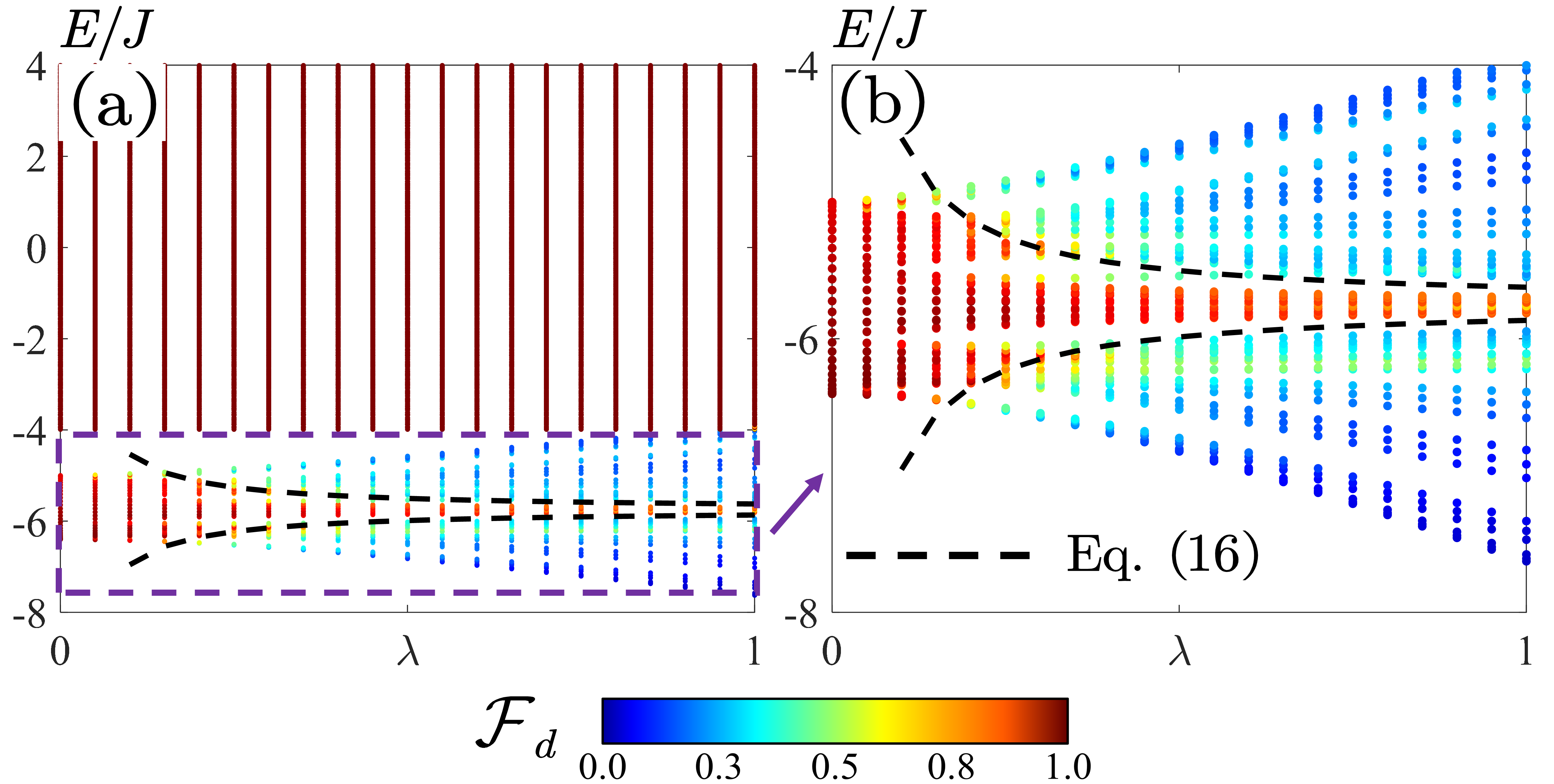}  
	\caption{(a) The $ \mathcal{F}_d $ of different doublon and scattering states in the two-photon subspace as a function of the quasiperiodic potential strength $\lambda$. In the scattering states, all states are extended, while MEs appear in the bound states. (b) An enlarged view of the doublon states. Since our study focuses on doublon states, subsequent figures only display the energy bands of the doublon states. The model parameters are $ \omega_c = 0 $, $ L = F_{18} = 2584 $, $ \kappa = 2 $, and $ U = -5J $. The black dashed lines represent the analytical solutions for MEs as described in Eq.~\eqref{eq14}.} 
	\label{fig4m} 
\end{figure}

For generality, we set $ \lambda = 1 $, $ \theta = 0 $, and choose $ \omega = (\sqrt{5} - 1)/2 $, a value that can be approximated using the Fibonacci numbers $ F_n $: $ \omega = \lim_{n \to \infty} F_{n-1}/F_n $, where $ F_n $ is defined by $ F_{n+1} = F_n + F_{n-1} $, with $ F_0 = F_1 = 1 $. To ensure periodic boundary conditions when numerically diagonalizing the tight-binding model defined in Eq.~\eqref{H_waveguide}, we set the system size to $ L = F_n $ and use the rational approximation $ \omega = F_{n-1}/{F_n} $.

Exact analytical solutions for MEs in 1D quasiperiodic mosaic models have been derived using Avila’s global theory to compute the Lyapunov exponent~\cite{Wang12020}. These solutions provide valuable insights into the localized and extended nature of all states within the energy spectrum. In our model, these findings are extended to the two-photon subspace by replacing the hopping constant in Ref.~\cite{Wang12020} with the effective hopping amplitude $ J_{\text{eff}} $ and incorporating the zero-point energy of the doublon band by setting $ K = \pi / 2 $ in Eq.~\eqref{eqEB}:

\begin{equation}
	E_0 = 2\omega_c - \sqrt{U^2 + 8J^2}.
\end{equation}

For $ \kappa = 2 $, the analytical expressions for the two MEs are given by:

\begin{equation}
	E_{c}^{(2)} = E_0 \pm \frac{J_{\mathrm{eff}}}{\lambda} = 2\omega_c - \sqrt{U^2 + 8J^2} \pm \frac{J_{\mathrm{eff}}}{\lambda}.
	\label{eq14}
\end{equation}

For $ \kappa = 3 $, the MEs are given by:
\begin{align}
	E_{c}^{(3)} &= E_0 \pm J_{\mathrm{eff}} \sqrt{1 \pm \frac{J_{\mathrm{eff}}}{\lambda}} \notag \\
	&= 2\omega_c - \sqrt{U^2 + 8J^2} \pm J_{\mathrm{eff}} \sqrt{1 \pm \frac{J_{\mathrm{eff}}}{\lambda}}.
	\label{eq15}
\end{align}

\begin{figure}
	\includegraphics[width=\linewidth]{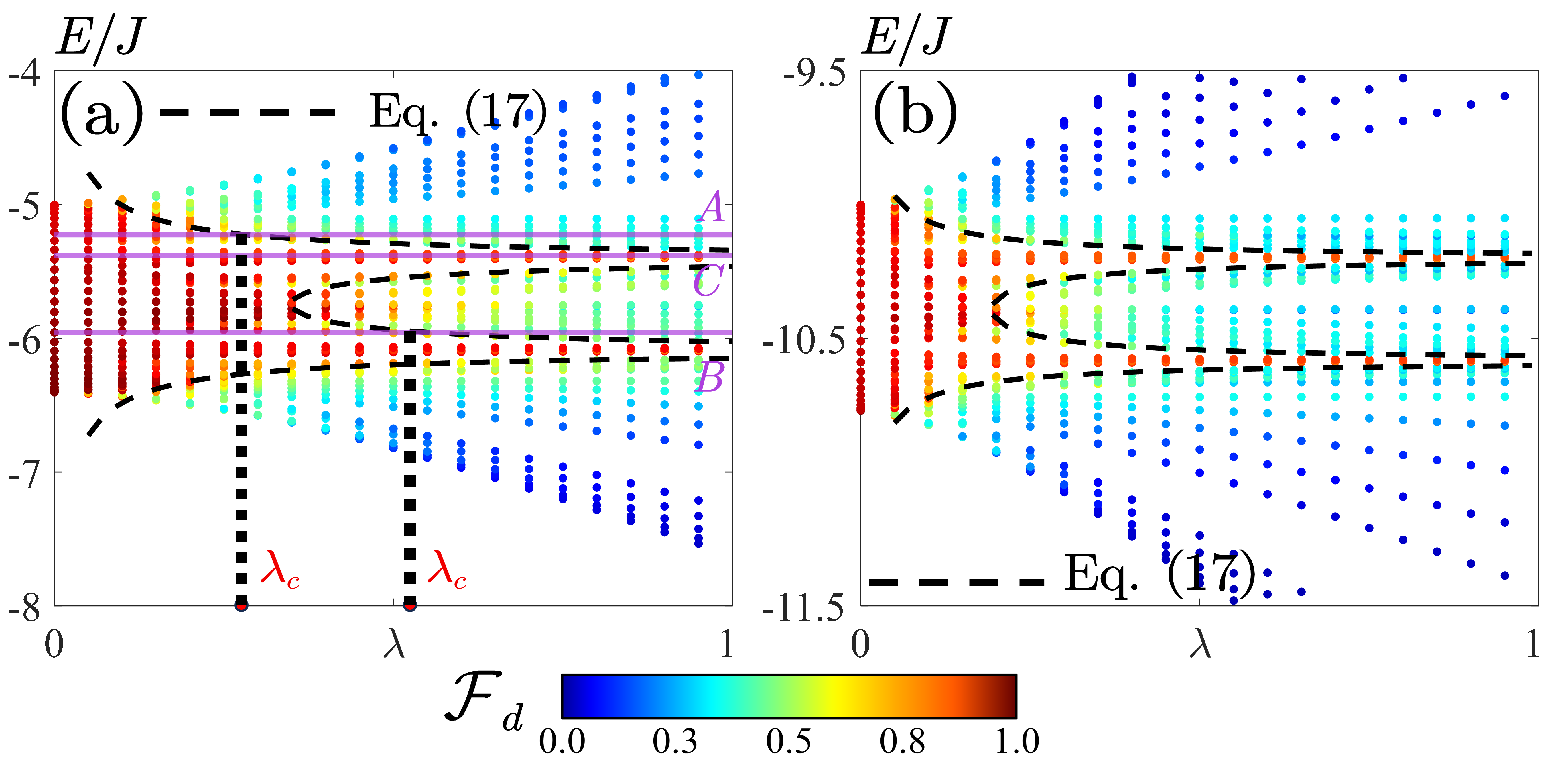}
	\caption{(a) and (b) illustrate the doublon states and MEs for different values of $U$. In panel (a), $U = -5J$, while in panel (b), $U = -10J$. As $U$ increases, the analytical solutions for MEs align more closely with the numerical solutions of the model. Model parameters are $\omega_c = 0$, $L = F_{18} = 2584$, and $\kappa = 3$. The black dashed lines represent the analytical solutions for MEs as described in Eq.~\eqref{eq15}. In panel (a), the three distinct purple solid lines, labeled as A, B, and C, represent the different eigenfrequencies of the two emitters, as defined in Eq.~\eqref{H_total}.} 
	\label{fig5m} 
\end{figure}

Different types of states can be characterized by their fractal dimension, denoted as $ \mathcal{F}_d $. For an arbitrary state $\ket{\psi_m} = \sum_{j=1}^{L} u_{m,j} D_j^\dagger \ket{\text{vac}}$, the $ \mathcal{F}_d $ is defined as
\begin{equation}
	\mathcal{F}_d = -\lim_{L \to \infty} \frac{\ln (\text{IPR})}{\ln L},
\end{equation}
where $\text{IPR} = \sum_{j} |u_{m,j}|^4$ is the inverse participation ratio. It is known that $ \mathcal{F}_d \to 1 $ for extended states and $ \mathcal{F}_d \to 0 $ for localized states.

In Fig.~\ref{fig4m} and Fig.~\ref{fig5m}, We plot energy eigenvalues and the $ \mathcal{F}_d $ of the corresponding eigenstates as a function of potential strength $ \lambda $. The energy band within the range $[-4J, 4J]$ corresponds to the scattering band with a bandwidth of $8J$, where $ \mathcal{F}_d $ values equal one, indicating that the wavefunctions are extended states. Just below the scattering band lies the doublon band, with the black dashed lines representing the analytical results for the MEs given by Eq.~\eqref{eq14} and Eq.~\eqref{eq15}. As predicted by our analytical results, the $ \mathcal{F}_d $ decreases from one to zero as the energies cross the black dashed lines, indicating a gradual shift in the wavefunctions of the doublon states from extended to localized states. This behavior is illustrated by the two purple solid lines, A and B, in Fig.~\ref{fig5m}(a). Notably, for $\kappa = 2$ and $\kappa = 3$, extended states persist at the center of the doublon band even when $ U \gg J $, as depicted by the purple solid line C in Fig.~\ref{fig5m}(a). In these cases, the hopping of the bound pair is akin to the previously mentioned second-order process. Consequently, as $U$ increases, the accuracy of $J_{\text{eff}}$ improves, resulting in a better alignment between the analytical and numerical results, as shown in Fig.~\ref{fig5m}. Since our primary focus is on the doublon band, we omit the scattering band in the subsequent figures, which will exclusively present the doublon band.

\begin{figure}
	\includegraphics[width=\linewidth]{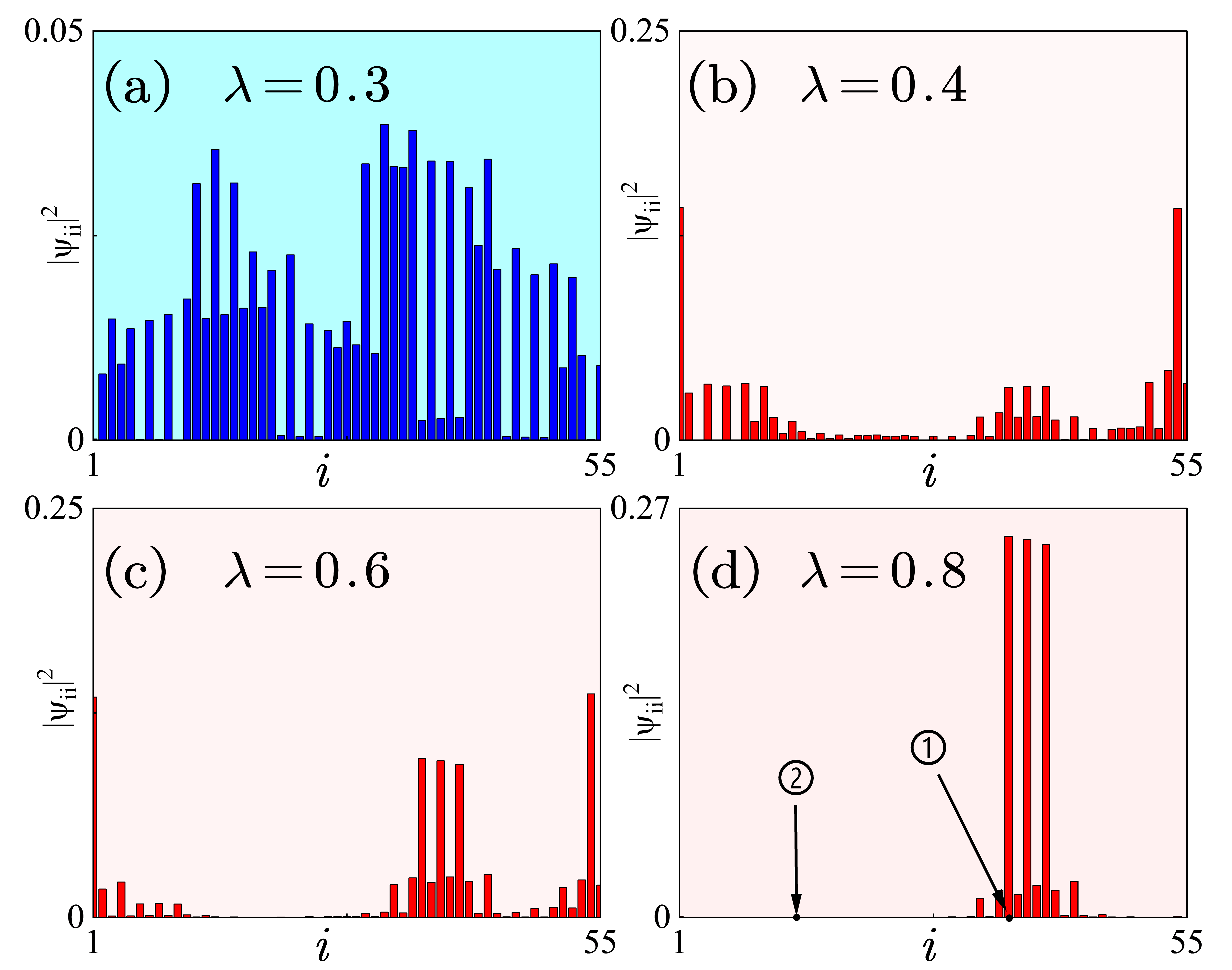} 
	\caption{The spatial distribution of the system's wavefunction along the diagonal line is illustrated here. We consider a chain length of $ L = F_{10} = 55 $, with parameters $ 2\omega_e = -5.77J $ and $ \kappa = 3 $. The critical value $\lambda_c = 0.35 $ is calculated from Eq.~\eqref{eq15}. Panel (a) shows that when $\lambda < \lambda_c $, the system is entirely in the extended state. Panels (b), (c), and (d) for $ \lambda > \lambda_c $, the system enters the localized state, and the degree of localization increases as $ \lambda $ becomes larger. In panel~(d), the lines labeled \textcircled{1} and \textcircled{2} correspond to the coupling of the two photons at different positions within the waveguide.} 
	\label{fig6m}  
\end{figure}

We can further determine the system's MEs by observing the spatial distribution of the wavefunction. We choose the waveguide chain length to be $ L = F_{10} = 55 $. From Eq.~\eqref{eq15}, we find that for the parameters $\kappa = 3$ and $2\omega_e = -5.77$, the system exhibits a critical value $\lambda_c = 0.35$. When $\lambda < 0.35$ (e.g., $\lambda = 0.3$), the system's wave function is in the extended state (see Fig.~\ref{fig6m}(a)). However, for $\lambda > 0.35$ (e.g., $\lambda = 0.4, 0.6, 0.8$), the system's wave function transitions to a localized state, with the degree of localization increasing as $\lambda$ is further increased (see Figs.~\ref{fig6m}(b)(c)(d)).

\begin{figure*}
	\includegraphics[width=\textwidth]{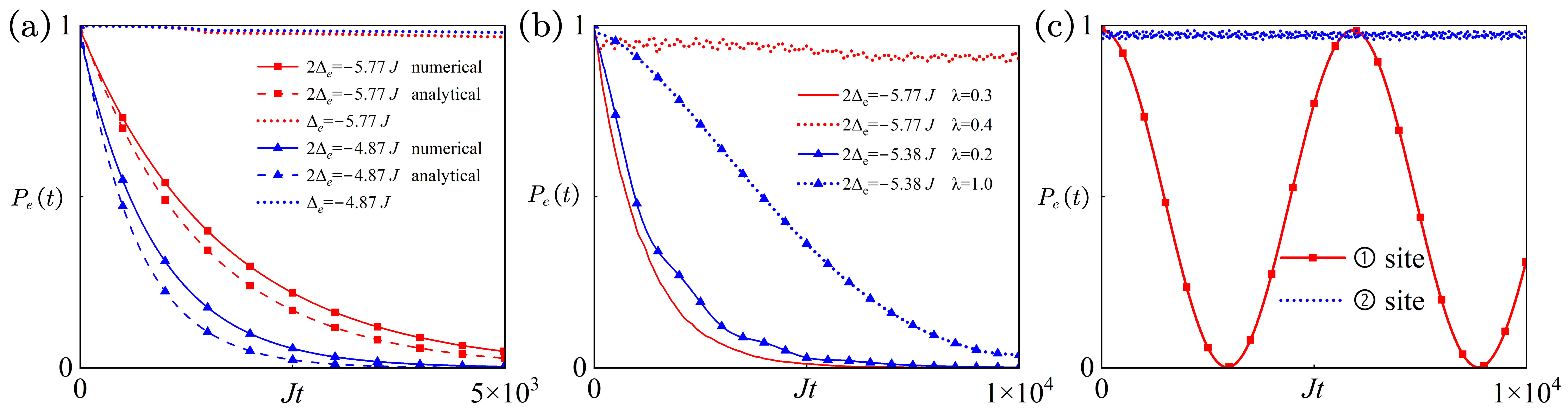}
	\caption{(a) For $\lambda = 0$, the curves depict the coupling of single and two emitters with the waveguide doublon band. The dotted line represents the single emitter case, while the solid line represents the two emitters case. The dotted line corresponds to the analytical result from Ref.~\cite{Wang22020}. (b) For $\lambda \neq 0$, the curves illustrate the two emitters coupling with the waveguide doublon band. The solid line corresponds to the case where $\lambda < \lambda_c$ (extended states), while the dashed line represents the case where $\lambda > \lambda_c$ (localized states). (c) The coupling of two emitters with the waveguide at different positions and the energy decay curves of the two-photon system are shown. \textcircled{1} and \textcircled{2} correspond to the different coupling positions depicted in Fig.~\ref{fig6m}(d). The parameters used are $ U = -5J $, $\kappa = 3$, and $g = 0.1J$.}
	\label{fig7m}
\end{figure*}

\section{The transition from Markovian to non-Markovian dynamics}

In this section, we investigate the transition from Markovian to non-Markovian dynamics, elucidating how the nature of the waveguide's eigenstates governs the emitter's energy decay. We begin by analyzing the emitter dynamics within the two-photon subspace, investigating how two emitters interact with various eigenmodes of the photonic waveguide. This analysis reveals the distinct behaviors that emerge when emitters couple to extended versus localized states. Following this, we present numerical simulations aimed at fitting the MEs predicted by our analytical model. These simulations not only validate our theoretical predictions but also provide comprehensive insights into the interplay between MEs and quantum dynamics in structured waveguide systems.

\subsection{Supercorrelated decay in two-photon subspace}

Through the above analysis, we have gained a clear understanding of the waveguide's modes and characteristics. Building upon this foundation, we consider the coupling of two two-level emitters to the waveguide, as described by the Hamiltonian in Eq.~\eqref{H_total}~\cite{Wang22020,Talukdar2022}. For convenience, we label $\Delta_e = \omega_c-\omega_e$. 

We set the frequency of the emitter $\Delta_e$ significantly detuned from the single-photon scattering state, i.e., $\Delta_e<-2J$. The dynamics of the emitter is totally suppressed, resulting in only a small residual decay, as shown by the red and blue dotted lines in Fig.\ref{fig7m}(a). However, due to the photon-photon interaction, the doublon state emerges within the two-photon subspace. When the total frequency of two emitters lies in doublon state, both emitter can cooperatively excite the doublon state and emit photons. Moreover, as discussed earlier, the doublon state consists of two strongly correlated (bound) photons. Therefore, during the emission process, the spatial separation between the two emitters is required smaller than the correlation length $L_c$, i.e., $|r_e|<L_c$, to ensure a nonzero overlap with the doublon wavefunction. The detail discussion about $r_e$ has been discussed in Ref.~\cite{Wang22020,Talukdar2022,Wangxin2024}. For our proposal, we assume $r_e=0$, where two emitters are coupled to the same point, allowing us to isolate and  study the effect of MEs on the dynamics of the emitters

Under these conditions, energy matching and nonzero wavefunction overlap, the supercorrelated radiation emerges, where \textit{the two excited emitters jointly emit photons and excite the doublon states.} The population of the state $|e_1,e_2\rangle $, two emitters being excited state simultaneously, $P_e(t)$ exhibits an exponential decay pattern characteristic of supercorrelated emission, as shown by the solid red (blue) curves in Fig.~\ref{fig7m}(a). Using the Wigner-Weisskopf approximation, the analytical decay rate is derives as:
\begin{equation}
\Gamma = \frac{2g^4}{J^3} \tilde{\rho}(K_0) f_{K_0}^2(n_1, n_2),
\label{decay_rate}
\end{equation}
For details on the derivation, we refer interested readers to Ref.~\cite{Wang22020}. Here, $ \tilde{\rho}(K) = J/v_g(K) $ represents the normalized density of bound two-photon states, and $ v_g(K) $ is the group velocity of the emitted photons, which depends on the eigenfrequency of the doublon states. Specifically, the group velocity at $ K_0 $ is
\begin{equation}
	v_g(K_0) = \left. \frac{\partial E_{\pm}^{B}}{\partial K} \right|_{K=K_0} = \frac{4J^2 \sin(K_0)}{\sqrt{U^2 + 16J^2 \cos^2(K_0/2)}}.
\end{equation}
The momentum $ K_0 $ is determined by the resonance condition $ 2\Delta_e = E_{\pm}^{B} $. Additionally, $ f_K(n_1, n_2) $ depends only on the relative distance between the two emitters coupling positions.

Compared to standard collective effects, i.e., super- and sub-radiance, resulting from the mutual influence of the emitters' radiation fields~\cite{Dicke1954,John1995,Roy2017,Shen2007,Sheremet2023,Gross1982}, supercorrelated radiance is a distinct multi-photon nonlinear collective phenomenon driven by the photon-photon interaction. In this regime, two excited emitters collectively and simultaneously radiate photons, described by the jointly quantum jump operator $S^{\pm}=\sigma^{\pm}_1\sigma^{\pm}_2$~\cite{Wangxin2024}. Furthermore, due to the excitation doublon state, the radiation photon field exhibits strongly spatial correlations, with the two photons bunched together and propagating along the waveguide as a single quasiparticle.

When $ \lambda = 0 $, the system reduces to the case described in Ref.~\cite{Wang22020,Talukdar2022,Wangxin2024}. In this situation, the wavefunction exhibits an extended behavior ($\mathcal{F}_d=1 $), spreading uniformly across the entire waveguide. The population dynamics of the emitters shows an exponentially supercorrelated decay pattern, as indicated by the solid red (blue) curves in Fig.~\ref{fig7m}(a). However, when $\lambda \ne 0$, the nonlinear potential is spatially modulated, introducing MEs into the doublon state. As $\lambda$ increases, the doublon wavefunction undergoes a  transition from an extended state $\mathcal{F}_d \simeq 1$ to a localized state $\mathcal{F}_d \simeq 1$. The critical value of $\lambda$ for this transition is denoted as $\lambda_c$. For $\lambda<\lambda_c$, the dynamics resemble the non-modulating situation ($\lambda=0$). The doublon wavefunction remains extended, and diffuse in the entire waveguide, enabling the emitters to emit the doublon state, regardless of their coupling position. However, when $\lambda>\lambda_c$, a novel phenomenon emerges, different from the scenarios described in Ref.~\cite{Wang22020,Talukdar2022,Wangxin2024}. Affected by the quasi-periodic modulation $\omega = F_{n-1}/F_{n}= 0.618$ (the golden ratio), the doublon wavefucntion becomes increasingly localized $\mathcal{F}_d\simeq 0$. In that scenario, the wavefunction is confined to specific regions, with negligible amplitude elsewhere. This implies the two photons are trapped within these regions and cannot freely propagate along the waveguide. Further, we consider the dynamics of emitters, which exhibits two interesting cases. First, when two emitters couple to the position 1 in Fig.~\ref{fig6m}(d), where the wavefunction is highly localized, the emitters can excite the doublon state. However, due to the localized, the doublon state cannot transport, and further is subsequently reabsorbed by the emitters, resulting in the Rabi oscillations. Conversely, when two emitters couple to position 2, where the wavefunction has negligible amplitude, the emitters are frozen in there excited state. 

These two scenarios can be understood through the Fermi Golden Rule: the transition rate depends on the overlap between the two-photon doublon state and the emitter state $|e_1,e_2\rangle$ (i.e., $|n_1,n_2\rangle$). When the emitters are located within regions where the doublon wavefunction has significant amplitude, the overlap is nonzero, leading to a nonzero transition rate. Conversely, if the emitters are positioned outside the support of the doublon wavefunction, the overlap vanishes, and no transitions occur. Compared to the traditional linear waveguide $\lambda=0$~\cite{Wang22020,Talukdar2022,Wangxin2024}, where the supercorrelated dynamics are primarily determined by the relative distance between two emitters, i.e., $r_e=|n_1-n_2|$, the situation becomes more complex when the nonlinear potential is modulated. In this case, the spatial position of the emitters [$x_c=(n_1+n_2)/2$] also plays a crucial role in the dynamics.

For $ \lambda \neq 0 $, the system exhibits disordered behavior, and consequently, only numerical solutions can be provided for the energy decay curves of the emitters. As shown in Fig.~\ref{fig7m}(b), using Eq.~\eqref{eq14} and Eq.~\eqref{eq15}, we find that for $2\Delta_e = -5.77$, $\lambda_c = 0.35$. The solid (dotted) lines in the figure correspond to the cases of $\lambda < \lambda_c$ and $\lambda > \lambda_c$, respectively, which is consistent with our analysis. Additionally, we have verified that the states remain extended for all values of $\lambda$, as indicated by the purple solid line C in Fig.~\ref{fig5m}(a). Moreover, the energy decay curves of these states consistently exhibit exponentially supercorrelated decay, as shown by the blue lines in Fig.~\ref{fig7m}(b).

\begin{figure*}[t]
	\includegraphics[width=\textwidth]{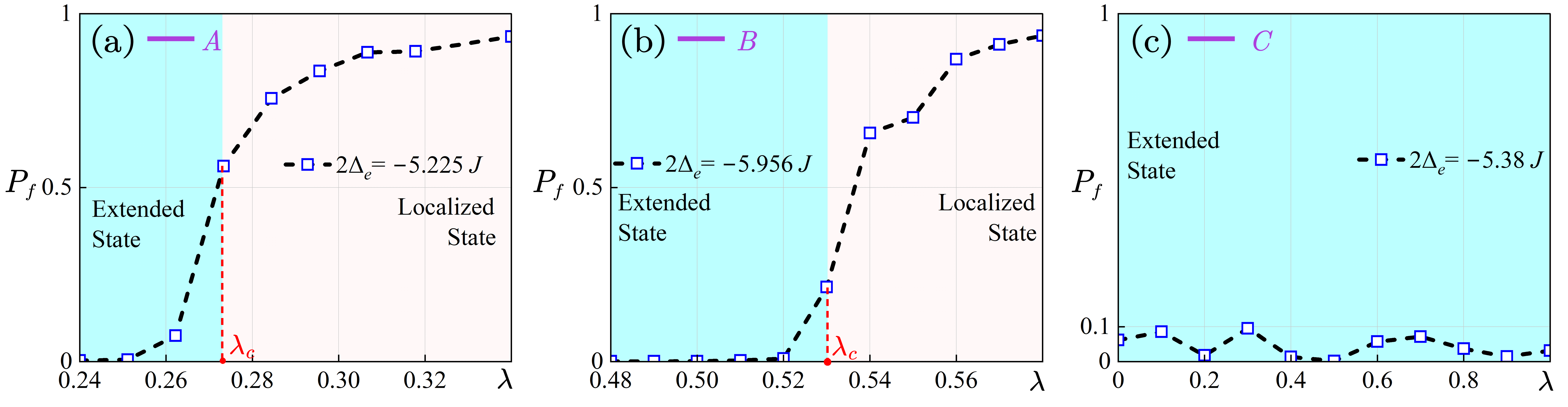}
	\caption{The variation of the final energy values $P_f$ of the energy decay curves for two emitters coupling with the doublon band at different frequencies is shown, with $U = -5J$, $g = 0.1J$, and $t = 2 \times 10^4 \, \text{s}$. Panels (a), (b), and (c) correspond to the three purple solid lines labeled A, B, and C in Fig.~\ref{fig5m}(a). In panels (a) and (b), the point where the $P_f$ begins to change corresponds to $\lambda_c$, which is determined through numerical simulations and marked by the red line in the figure. In panel (c), no $\lambda_c$ is observed, indicating that the system's eigenstates remain in the extended state.} 
	\label{fig8m} 
\end{figure*}

\subsection{Numerical simulation around Mobility Edges}

Through an in-depth analysis of emitters dynamics, we can determine the type of coupling between the emitters and the waveguide—whether they are coupled to an extended state or a localized state—based on the energy decay curves. Furthermore, numerical simulations enable us to more precisely identify the presence of MEs. We can leverage this characteristic to obtain MEs under various parameters, enabling us to fit the overall curve of the MEs.

The procedure is as follows: we select a specific $2\Delta_e$ (as shown by the purple solid line in the Fig.~\ref{fig5m}(a), ensuring that $2\Delta_e < -4J$, placing it within the bound state. As the amplitude of the quasiperiodic potential $\lambda$ gradually increases, the system’s wave function transitions from an extended state to a localized state, crossing the critical point $\lambda_c$. Correspondingly, the energy curves of the emitters shifts from exponentially supercorrelated decay to a trapped state (as indicated by the transition from the solid line to dashed line in Fig.~\ref{fig7m}(b)). The point at which the energy decay curves begins to change marks the critical value $ \lambda_c $, which also corresponds to the transition from Markovian to non-Markovian dynamics in the system. By repeating this procedure with different choices of $2\Delta_e$, We can obtain the results of the numerical through this characteristic.

To further analyze the system's dynamics, we consider emitters initially in the excited state, i.e., $P_i = |C_e(t=0)|^2 = 1$. After long-time evolution, the final state of the emitters can be expressed as $P_f = |C_e(t \to \infty)|^2$, representing the probability of the emitters remaining in the excited state. By examining this final state $P_f$, we can assess changes in the system's eigenstates through the energy curves, as shown in Fig.~\ref{fig8m}(a)(b)(c). These curves are plotted as functions of $\lambda$ at a specific emission energy of $2\Delta_e$, with the time set to $t = 2 \times 10^4 \, \text{s}$. In Fig.~\ref{fig8m}(a)(b), When $\lambda < \lambda_c$, the final state $P_f$ approaches zero, indicating that the energy of the emitters has fully diffused into the lattice. As $\lambda$ increases and exceeds $\lambda_c$, the $P_f$ gradually rises, indicating that the energy of the emitters starts to be trapped by the high potential well. This $\lambda_c$ is the result obtained from our numerical simulation. In Fig.~\ref{fig8m}(c) regardless of how $\lambda$ changes, the $P_f$ consistently approaches zero, indicating that the system's eigenstates are always extended states.

\begin{figure}
	\includegraphics[width=\linewidth]{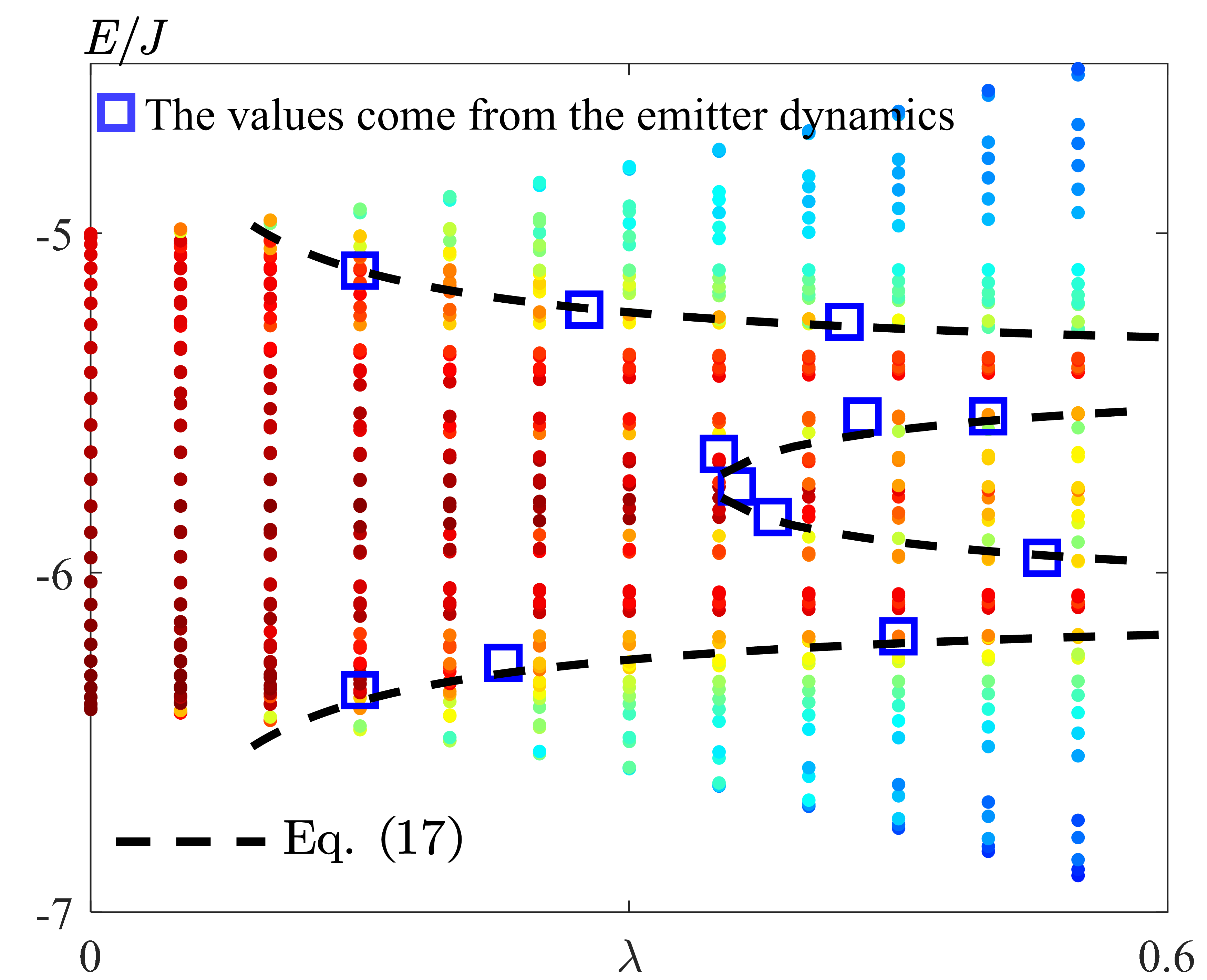} 
	\caption{Comparison between numerical and analytical solutions in the two-photon subspace. The blue squares represent the numerical results from the emitter dynamics, while the black dashed line shows the analytical solution in Eq.~\eqref{eq15}, with the same parameters used in Fig.~\ref{fig5m}(a).} 
	\label{fig9m} 
\end{figure}

In Fig.~\ref{fig9m}, we compare the numerically obtained MEs derived from the energy decay curves of the emitters with the analytical MEs given by Eq.~\eqref{eq15}. The figure illustrates a remarkable agreement between the numerically fitted curve and the analytical solution, validating the robustness of our analytical framework. This close alignment not only reinforces the reliability of our model but also highlights the intricate interplay between the bound states and the MEs in the two-photon subspace. Such findings pave the way for future investigations into the dynamics of quantum systems influenced by similar quasiperiodic potentials.

\section{Experimental realization with circuit-QED}

\begin{figure}
	\includegraphics[width=\linewidth]{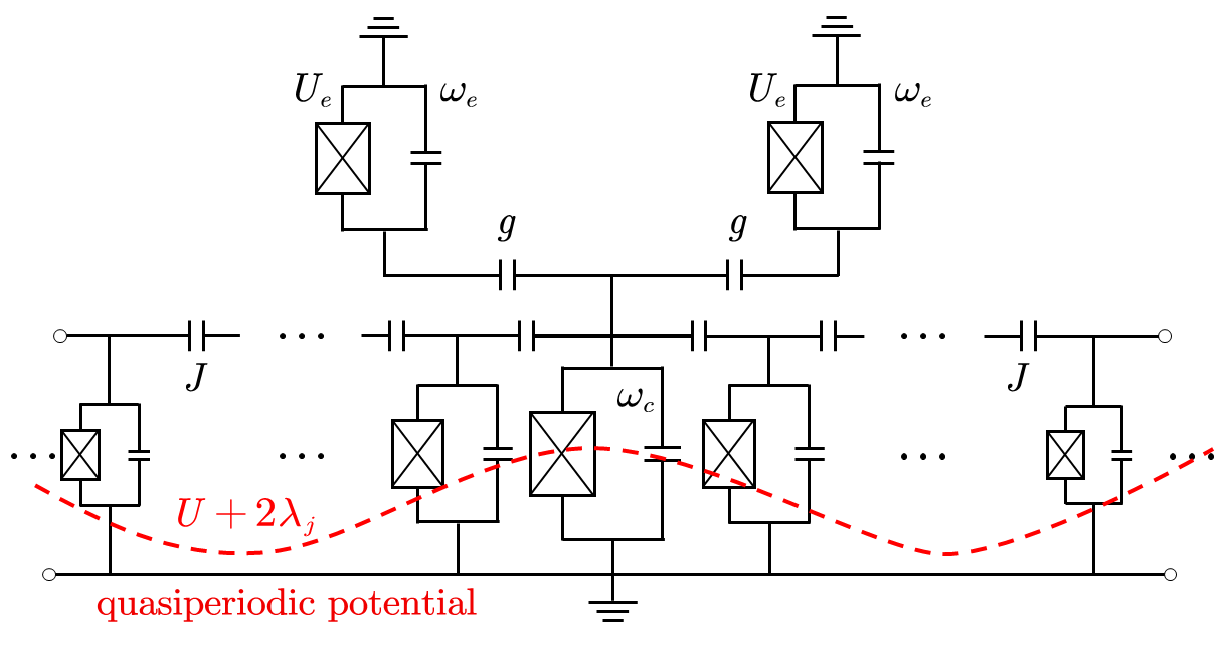}
	\caption{Schematic of the superconducting circuit implementing the Hamiltonian in Eq.~\eqref{H_total}. Each site of the one-dimensional waveguide is realized by a transmon qubit, which consists of a Josephson junction and capacitor pads. The size of the transmons, indicated by the red dashed line, illustrates the quasiperiodic modulation of the anharmonicity, corresponding to the nonlinear potential $U$ in Eq.~\eqref{eq:lambda}. Neighboring transmons are capacitively coupled to enable photon hopping between adjacent sites. Two identical emitter transmons above the transmon array are capacitively coupled to the same lattice site of the waveguide, facilitating local emitter–photon interactions with strength $g$.
	}
	\label{fig10m}
\end{figure}

\begin{figure}
	\includegraphics[width=\linewidth]{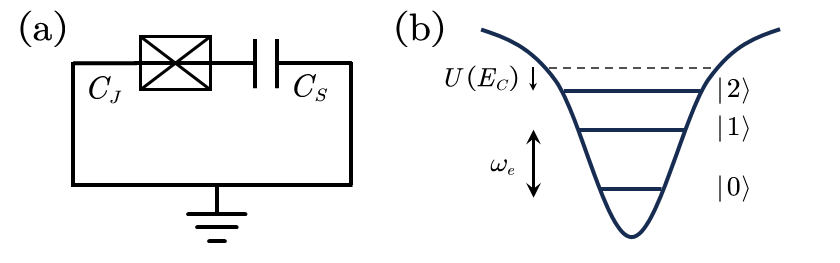} 
	\caption{(a) Schematic of a Josephson qubit circuit, with the junction capacitance $C_J$ and the shunt capacitance capacitance $C_S$. (b) The resulting energy spectrum features  non-equidistant energy levels, a consequence of the circuit's intrinsic anharmonicity $E_C$, i.e., the nonlinear potential $U$.} 
	\label{fig11m}
\end{figure}

To implement the Hamiltonian shown in Eq.~\eqref{H_total}, we map its physical parameters onto components of a superconducting circuit~\cite{Koch2007,Orell2019,Krantz2019,Blais2021,Carusotto2020,Dalmonte2015,Deng2016,Roushan2017,Ma2019,Ye2019,Wilkinson2020,Yanay2020,Fedorov2021,Kim2021,Mansikkam2022,Xiang2023}, thereby constructing an experimentally realizable quantum platform, as illustrated in Fig.~\ref{fig10m}. Each site in the one-dimensional waveguide is implemented using a transmon qubit, as shown in Fig.~\ref{fig11m}(a). The transmon qubit consists of two capacitor pads connected by a Josephson junction, forming an anharmonic oscillator. The Hamiltonian is written as ~\cite{Koch2007,Orell2019,Krantz2019,Blais2021}
\begin{gather}
H\simeq \omega _cb^{\dagger}b-\frac{E_C}{2}b^{\dagger}b^{\dagger}bb, \label{H_exp}
\end{gather}
where $\omega _c=\sqrt{8E_CE_J}-E_C$ is the fundamental transition frequency, $E_J$ is the Josephson energy, and $E_C = e^2 / (2C_\Sigma)$ is the charging energy. Here, $C_\Sigma = C_S + C_J$ denotes the total capacitance, which includes contribution from the shunt capacitance $C_S$ and the junction capacitance $C_J$. Due to the intrinsic anharmonicity $E_C$, the non-equidistant energy levels of a transmon are shown in Fig.~\ref{fig11m}(b). In our design, we set $\omega _c/2\pi \sim 5\,\mathrm{GHz}$, a typical value in superconducting circuit experiments.

The hopping term in the Hamiltonian is realized through capacitive coupling between adjacent transmons, allowing coherent photon tunneling between neighboring sites. By adjusting the capacitive coupling between adjacent lattice sites, we can set a fixed nearest-neighbor tunneling strength $J$. We choose $J/2\pi \sim 40\mathrm{MHz}$, which provides an effective support for the propagation of quantum states and ensures good experimental controllability. The nonlinear local potential $U$ arises from the intrinsic anharmonicity of the transmon. Compared Eq.~(\ref{BH}) and Eq.~(\ref{H_exp}), the photon-photon interaction is $U=E_C$. $E_C$ is determined by  the geometric and dielectric properties of the capacitor. Since these properties are fixed during fabrication, $U$ is also predetermined at this stage. In our implementation, we design transmons with a target interaction strength $U/2\pi \sim 200\,\mathrm{MHz}$, with $U\sim 5J$. To achieve quasiperiodic modulation of $U$, we propose fabricating transmons with site-dependent capacitor sizes or dielectric constants. This approach induces a spatially varying $E_C$, which in turn results in a modulated interaction strength $U$ across the array.

\begin{figure}
	\includegraphics[width=\linewidth]{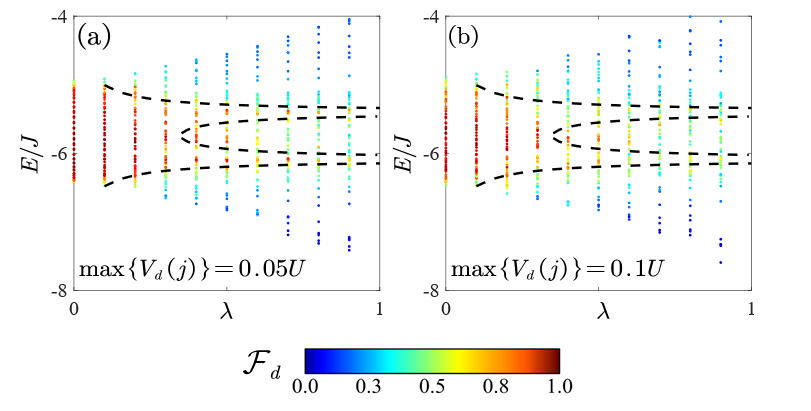} 
	\caption{The energy spectrum of the Hamiltonian Eq.~(\ref{H_disorder}) is shown for two cases: (a) the maximum disorder strength of $\max\{V_d(j)\} = 0.05U$, and (b) $\max\{V_d(j)\} = 0.1U$. All other parameters are the same as those in Fig.~\ref{fig5m}.} 
	\label{fig12m}
\end{figure}

Note that, in experimental realizations, $E_C$ cannot be precisely controlled during fabrication, leading to disorder in the quasiperiodic modulation of the nonlinear potential $U$ modulation. To account for this effect, we introduce an effective disorder term into the Hamiltonian as follows:
\begin{gather}
H = H_{w} + \sum_{j}V_{d}(j)a^{\dagger}_ja^{\dagger}_ja_ja_j, \label{H_disorder}
\end{gather}
where $V_{d}(j)$ represents the random disorder in the modulation potential. Figure~\ref{fig12m} shows the energy spectrum as a function of the modulation parameters for different $\max \{V_{d}(j)\}$. Even the $V_d(j)$ reaches up to $10\%$ of the nonlinear $U$, the MEs (the boundary of extended and localized state) still exist, although the states are perturbed. Furthermore, the interplay between Anderson localization (disorder) and quasiperiodic modulation in shaping the energy structure of the bath or the dynamics of the emitters remains an open question. This intriguing topic will be explored in future work~\cite{Segev2013}.
	
For the two two-level emitters depicted above the transmon array in Fig.~\ref{fig10m}, the transmon anharmonicity is set to $U_e/2\pi \sim 300 \mathrm{MHz}$. This setup effectively restricts the transmons to behave as two-level systems by suppressing transitions to higher energy levels~\cite{Orell2019}. Two transmons (emitters) are coupled at the same lattice site, and the coupling strength $g$ is controlled by adjusting the coupling capacitance between the transmon (emitter) and the waveguide site. This coupling enables us to observe interactions between the transmons and the photon modes in the superconducting circuit. We set the coupling strength $g/2\pi \sim 4\mathrm{MHz}$ ensuring that the interaction between the emitters and the photon modes meets our expectations and effectively induces quantum transitions. To quantify the energy relaxation process of the transmon, we use microwave pulses to excite the transmon from its ground state to the excited state, and we measure the probability of excitation decay over time. We select the emitter's emission frequency around $\Delta_e = (\omega_e - \omega_c)/2\pi \sim -100 \mathrm{MHz}$, which allows resonance between the emitter and different photon modes in the waveguide, enabling us to observe energy decay curves for various modes.

Moreover, the supercorrelated decay rate in Fig.~\ref{fig7m}(a) is on the 
order of $\Gamma / (2\pi) \sim {0.1}\mathrm{MHz}$. In contrast, typical 
intrinsic decay rates $\gamma$ and dephasing rates $\kappa$ in transmon 
arrays are significantly smaller. For instance: the intrinsic decay rate is 
approximately $\gamma / (2\pi) \simeq 15\,\mathrm{kHz}$, corresponding to a 
relaxation time $T_1 \sim 10\,\mathrm{\mu s}$; the dephasing rate is 
approximately $\kappa / (2\pi) \sim 3\,\mathrm{kHz}$, corresponding to 
coherence times $T_2 \sim 50\,\mathrm{\mu 
s}$~\cite{Kjaergaard2020,Place2021,Krinner2022}. Given these values, the 
supercorrelated decay rate in our proposal is significantly larger than both 
the intrinsic decay rate and the dephasing rate, i.e., $\Gamma \gg \{\gamma, 
\kappa\}$. This suggests that the supercorrelated decay dynamics should be 
observable within the time windows set by the dissipation and dephasing times 
of current circuit-QED platforms.

\section{Conclusion}

In this work, we examine the two-photon Bose-Hubbard chain, focusing on two-photon bound state (doublon) under varying nonlinear local potential strengths~\cite{Valiente2008,Winkler2006}. In the strong-interaction regime, doublon transitions simplify to an effective second-order process, where the doublon pair is treated as a single-photon tight-binding model with an effective hopping amplitude and Hamiltonian. We then introduce a one-dimensional quasiperiodic potential to modulate the nonlinear potential, transforming the system into a quasiperiodic mosaic lattice. Numerical diagonalization reveals MEs that separate localized and extended states. Extending previous work~\cite{Wang12020}, we analyze two-photon MEs by replacing the hopping constant with an effective hopping amplitude. The close agreement between analytical and numerical results provides a solid framework for studying nonlinear effects in two-photon systems.

Furthermore, we investigate the dynamical behavior of two emitters coupled to a critical quasiperiodic bath, revealing a transition from Markovian exponential decay to non-Markovian dynamics, which strongly depends on the quasiperiodic modulation. This transition also has recently been realized by tuning the disorder parameter~\cite{Monteiro2024}. Notably, in the non-Markovian regime, the dynamics depend not only on the modulation parameters but also on the coupling position of the emitters, ranging from coherent oscillations to complete freezing. The critical quasiperiodic potential plays a pivotal role in governing these transitions, highlighting the unique ability of quasicrystals to shape light-matter interactions through both parametric and spatially dependent dynamics.

Our findings deepen the understanding of quantum emitter dynamics in complex environments, and open new avenues for exploring how quasiperiodic structures influence quantum optical processes, which remains relatively underexplored. Moreover, we propose an experimental implementation using superconducting circuits, providing a scalable platform for observing these phenomena. This approach holds promise for practical applications, such as engineering tunable quantum systems with parametric- or position-dependent radiation control.

\section*{Acknowledgements}
X.W.~is supported by the National Natural Science Foundation of China (NSFC) 
(Grant No.~12174303).

\appendix

\section{Energy Bands and Wave Function Solutions in the Two-Photon Subspace}\label{appendix1}

A general two-photon eigenstate for the Bose–Hubbard model can be expressed as~\cite{Piil2007,Wang22020}:  
\begin{equation}
	\ket{\Psi} = \frac{1}{\sqrt{2}} \sum_{m,n} \Psi(m,n) a_m^\dagger a_n^\dagger \ket{\text{vac}},
	\label{eq6}
\end{equation}
where the factor $ 1/\sqrt{2} $ arises from the bosonic symmetry. The function $ \Psi(m, n) $ is symmetric and describes the probability amplitude for two photons localized at the position $ (m, n) $.

By introducing the center of mass coordinate $ X_c = (m+n)/2 $ and the relative coordinate $ r = m - n $, the eigenstate can be written in ansatz form as $ \Psi(m, n) = e^{i K X_c} \psi_K(r) $. Here, $K$ represents the momentum of the center of mass, and $\psi_K(r)$ denotes the wavefunction of the bound pair. Substituting $ H_{\text{BH}} $ and $ \Psi(m, n) $ into the Schrödinger equation, we obtain,
\begin{align}
	2\omega_c \psi_K(r) & - 2J \cos\left(\frac{K}{2}\right) [\psi_K(r+1) + \psi_K(r-1)] \nonumber \\
	& + U \delta_{r,0} \psi_K(r) = E \psi_K(r).
	\label{eq8}
\end{align}
Here, $ \delta_{r, 0} $ is the Kronecker delta function, which indicates that the nonlinear potential $ U $ is effective only when $ r = 0 $. This means that the nonlinear interaction acts only when two photons occupy the same site.

Next, using perturbation theory, we define the scattering potential as $ V(r) \equiv U \delta_{r, 0} $, and then Eq.~\eqref{eq8} becomes equivalent to:
\begin{equation}
	H_0 \psi_K(r) + V(r) \psi_K(r) = E \psi_K(r).
\end{equation}
Let us first consider the solution for the scattering band. Since the wavefunctions of the scattering states extend throughout the entire lattice, the two-photon scattering state can be approximated by the following trial solution:
\begin{equation}
	\psi_K^S(r) = A \cos(kr) + B \sin(kr).
\end{equation}
Substituting this into Eq.~\eqref{eq8}, we obtain the energy dispersion relation and wavefuntion for the scattering band:
\begin{equation}
	E_S = 2\omega_c - 4J \cos\left(\frac{K}{2}\right) \cos(k),
	\label{eq9}
\end{equation}
\begin{equation}
	\psi _{K}^{S}\left( r \right) =A\left( \cos\mathrm{(}kr)+\frac{U\sin\mathrm{(}kr)}{4J\sin \left( k \right) \cos \left( \frac{K}{2} \right)} \right). 
\end{equation}
As shown by the solid blue line in Fig.~\ref{fig2m}(a), these states are unbound and disperse freely throughout the entire lattice, allowing the particles to move independently. The bandwidth of the two-photon scattering spectrum is $E_S = 8J$, which is twice the width of the single-photon scattering spectrum. This can be interpreted as a linear superposition of two single-photon scattering states.

Next, we solve for the bound state. Here, the nonlinear potential is treated as a scattering potential, and the bound state solution is handled as a perturbative correction to the wavefunction. We define the scattering potential $V(r) \equiv U \delta_{r,0}$, simplifies Eq.~\eqref{eq9} to:
\begin{equation}
	H_0 \psi_K(r) + V(r) \psi_K(r) = E \psi_K(r).
\end{equation}
According to the Green's function method from scattering theory, in the absence of the nonlinear term, the unperturbed Green's function $ G_{K}^0(E, r) $ is determined by:
\begin{equation}
	(E - H_0) G_{K}^0(E, r) = \delta_{r,0}.
\end{equation}
To handle the delta function, we perform a Fourier transform, which converts the Green's function to momentum space:
\begin{equation}
	G_{K}^0(E, r) = \frac{1}{2\pi} \int dq \, G_{K}^0(E, q) e^{iqr}.
\end{equation}
The Green's function in momentum space then satisfies the algebraic equation:
\begin{equation}
	(E - H_0) G_{K}^0(E, q) = 1.
\end{equation}
It is straightforward to verify that $ e^{iqr} $ is an eigenfunction of $ H_0 $, where:
\begin{equation}
	H_0 e^{iqr} = (2\omega_c-4J \cos\left(\frac{K}{2}\right) \cos(q)) e^{iqr}.
\end{equation}
Thus, the Green's function in momentum space becomes:
\begin{equation}
	G_{K}^0(E, q) = \frac{1}{E + 4J \cos\left(\frac{K}{2}\right) \cos(q)-2\omega_c}.
\end{equation}
With the explicit form of the Green's function, the bound state solution can be expressed in the Lippmann-Schwinger form as:
\begin{equation}
	\psi_K(r) = \psi_K^S(r) + \int dr' \, G_{K}^0(E, r - r') V(r') \psi_K(r').
\end{equation}
Here, $ \psi_K^S(r) $ is the scattering state solution in the absence of the nonlinear potential, and the second term represents the bound state solution $ \psi_K^B(r) $ induced by the nonlinear potential. After simplifications, we have:
\begin{align}
	\psi_K^B(r) &= \int dr' \, G_{K}^0(E, r - r') U \delta_{r',0} \psi_K(r') 
	\notag \\
	&= G_{K}^0(E, r) U \psi_K(r=0).
\end{align}
Thus, the Lippmann-Schwinger equation simplifies to:
\begin{equation}
	\psi_K(r) = \psi_K^S(r) + G_{K}^0(E, r) U \psi_K(r=0).
\end{equation}
When the two photons are located at the same lattice site, the bound state coefficient tends to infinity, indicating a pure bound state solution is obtained. Setting $ r = 0 $ yields:
\begin{equation}
	\left[ 1 - G_{K}^0(E, r=0) U \right] \psi_K(r=0) = \psi_K^S(r=0).
\end{equation}
Setting the determinant condition:
\begin{equation}
	\text{det}\left[ 1 - G_{K}^0(E, r=0) U \right] = 0,
\end{equation}
we find:
\begin{align}
	\frac{1}{U} &= G_{K}^0(E, r=0) 
	\notag \\
	&= \int dq \, \frac{1}{E + 4J \cos\left(\frac{K}{2}\right) \cos(q)-2\omega_c}.
\end{align}

Finally, we obtain the two-photon bound state energy band:
\begin{equation}
	E_{\pm}^{B} = 2 \omega_c \pm \sqrt{U^2 + \left[ 4J \cos\left( \frac{K}{2} \right) \right]^2}.
\end{equation}

To proceed with solving the bound-state eigenfunction for the two-photon system, we follow the derivation from the previous sections and obtain the expression:
\begin{equation}
	\psi_K^B (r) = G_K^0 (E, r) U \psi_K (r=0).
\end{equation}
For simplicity, we define $ J_K \equiv J \cos\left( \frac{K}{2} \right) $. Using this definition, the bound-state wave function can be expressed as:
\begin{equation}
	\psi_K^B (r) = \psi_K^B (0) \left( \frac{\sqrt{U^2 + 16 J_K^2} - |U|}{4 J_K} \right)^r.
\end{equation}
This expression accounts for both attractive and repulsive nonlinear potentials, where $ \psi_K^B (0) $ is the normalization factor. Therefore, the final two-photon bound state solution is a combination of an extended state in the center-of-mass direction and a localized state in the relative position direction:
\begin{equation}
	|\Psi\rangle = \frac{1}{\sqrt{2}} \sum_{m,n}^N e^{iK (m+n)/2} \psi_K^B (m-n) a_m^\dagger a_n^\dagger |\mathrm{vac}\rangle.
\end{equation}

%


\end{document}